\documentstyle[11pt,newpasp,twoside, epsf]{article}
\def\edcomment#1{\iffalse\marginpar{\raggedright\sl#1\/}\else\relax\fi}
\reversemarginpar

\begin{document}
\title{Dissipation of non-linear circularly polarized Alfv\'{e}n waves}
\author{Rim Turkmani}
\author{Ulf Torkelsson}
\affil{
Department of Astronomy and Astrophysics \\
Chalmers University of Technology/G\"oteborg University \\
S-412 96 Gothenburg, Sweden}

\begin{abstract}
We study propagating Alfv\'{e}n waves by solving the time-dependent 
equations of
magnetohydrodynamics (MHD) in one dimension numerically. In a homogeneous 
medium
the circularly polarized Alfv\'{e}n wave is an exact solution of the ideal
MHD equations, and therefore it does not suffer from any dissipation.  A
high-amplitude linearly polarized Alfv\'{e}n wave, on the other hand, 
steepens and form current 
sheets, in which the
Poynting flux is lost.
In a stratified medium, however, a high-amplitude circularly polarized 
Alfv\'{e}n wave can also lose a significant fraction of its Poynting flux.

\end{abstract}

\section{Introduction}

The acceleration of the fast solar wind is a long-standing problem in solar 
physics.
It is likely that Alfv\'{e}n waves play a significant role in this process
(e.g. Leer et al. 1982).
They can propagate over long distance, which allows them to
reach the outer corona, because to lowest order they are incompressible and 
do not dissipate.
However dissipative damping is required at some point to avoid too high wind 
velocities (e.g. Holzer et al. 1983).
A closer examination shows though that a linearly polarized Alfv\'{e}n wave 
is compressible to second order, since the magnetic pressure,
$B_1^2 / (2 \mu_0)$, varies
with half the wave length of the Alfv\'{e}n wave itself
(Alfv\'{e}n \& F\"{a}lthammar 1963). In a circularly 
polarized Alfv\'{e}n
wave, on the other hand, the magnetic pressure is constant along the wave,
which is the physical reason why the circularly polarized Alfv\'{e}n 
wave
in a homogeneous medium is an exact solution to the nonlinear MHD 
equations.

Several groups have studied different mechanism through which Alfv\'en waves
can dissipate.  In two dimensions phase mixing due to
a transverse gradient in the phase velocity (e.g. Heyvaerts \& Prist 1983)
can lead to a strong damping of the wave.  This mechanism is not available
in one dimension though, and one has then rather considered the nonlinear
coupling of the Alfv\'en wave with other modes (e.g. Wentzel (1973), for
circularly polarized Alfv\'en waves in particular see also Del Zanna et al.
(2001) ).  Cohen \& Kulsrud (1974) found in an approximate solution of the
nonlinear MHD equations that the linearly polarized Alfv\'en wave 
steepens and can form current sheets.  This result has later been confirmed
by numerical simulations by Boynton \& Torkelsson (1996) and
Ofman \& Davila (1997).

\section{Results}
In this paper we use the numerical code of Boynton \& Torkelsson (1996) (see 
also Torkelsson \& Boynton (1998)) to simulate propagating circularly polarized 
Alfv\'{e}n waves.
In the following we will consider both models with a constant background 
density, and stratified models.
For all the simulations the temperature is $10^6$ K and the period of the 
waves is 300 s. The isothermal speed of sound is $1.29 \, 10^5$ m\,s$^{-1}$.


\begin{figure}
\epsfxsize=13.0cm
\epsfbox{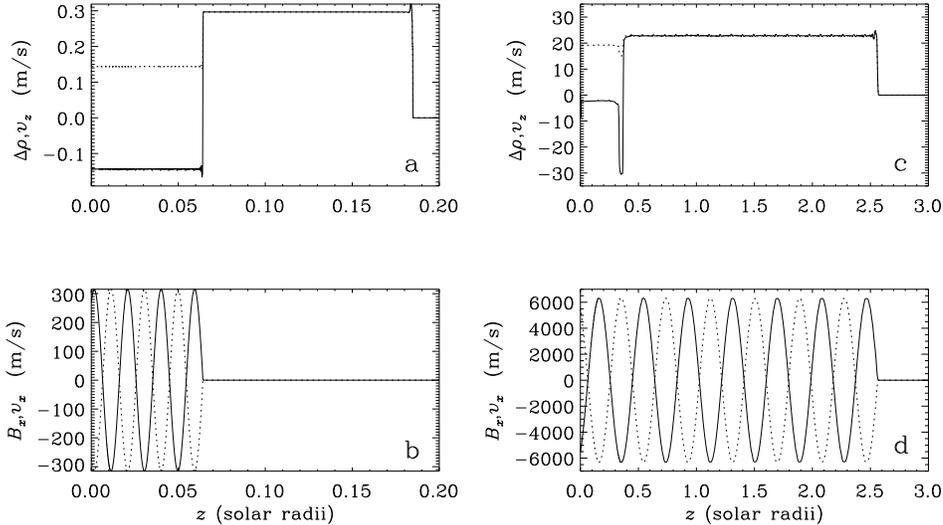}
\caption{Magnetohydrodynamics waves propagating through a
homogeneous medium and wave amplitude $(\frac{B_x}{B_z})_0 = 0.01$. 
In {\bf a, b}  $\beta_{\rm mag}=17$ and in {\bf c, d} $\beta_{\rm mag}=0.042$.
We plot in {\bf a} and {\bf c} $ \Delta \rho = \rho - \rho_0$ (solid line) 
and $v_z$ (dashed line)
as a function of distance, $z$, at 2000 s. $\Delta \rho$ has been converted
to velocity units by multiplying with $c_{\rm s}/{\rho_0}$ in {\bf a} and 
with $v_{\rm a}/{\rho_0}$
in {\bf c}, where $\rho_0$ is the background density.
In {\bf b} and {\bf d} we plot $B_x$ (solid line) and $v_x$ (dashed line) as 
a function of distance
$z$ at the same time. $B_x$ has been converted to velocity units by 
multiplying
with $1/{\sqrt{\mu_0 \rho_0}}$.}
\end{figure}


\subsection{Unstratified models}

For the unstratified models, we differentiate between two cases, when the
magnetic field is weak, that is $\beta_{\rm mag}=17$ and $c_{\rm s} > v_{\rm 
A}$, and when it is strong that is $\beta_{\rm mag}=0.042$ and $v_{\rm A} >c_{\rm 
s} $.
For both the runs $\rho=1 \, 10^{-14}$\,kg\,m$^{-3}$.
We compare the runs in Fig. 1.
If we compare our Fig 1a to Fig. 1 of Boynton \& Torkelsson (1996) we notice 
that in both cases there is an acoustic precursor moving at a speed close to 
$c_s$, but in their case the precursor is a sinusoidal oscillation in $\Delta \rho$ and 
$v_z$.
Apparently the gradual increase of the magnetic pressure at the front of the 
linearly polarized Alfv\'en wave drives an oscillatory wave, while the sharp 
discontinuity of the magnetic pressure at the front of the circularly
polarized wave works as a piston.
The amplitude of the acoustic precursor in Fig. 1a scales with
the square of the amplitude of the Alfv\'{e}n wave.

When $v_{\rm A} > c_{\rm s}$ (Fig. 1c and 1d)
a density enhancement propagates with the wave. $v_z$ and
$\Delta \rho$ are then related by $v_z=v_{\rm A} \frac{\Delta \rho}{\rho}$.
While the right edge of the enhancement coincides with the front of the
Alfv\'{e}n wave, the left edge propagates with
the velocity  $c_{\rm s}$.
In the leftmost part of Fig. 1c we see the effect of the non-linear
interaction between the acoustic wave and the Alfv\'en wave.

Boynton \& Torkelsson (1996) found that when $B_1$ approaches $B_z$ in
strength, the Alfv\'en wave starts to steepen into current sheets (Fig. 2a) 
which enhance
the dissipation of the wave. The circularly polarized Alfv\'{e}n wave does 
not
develop such current sheets (Fig. 2b).

\begin{figure}
\epsfxsize=6.0cm
\epsfbox{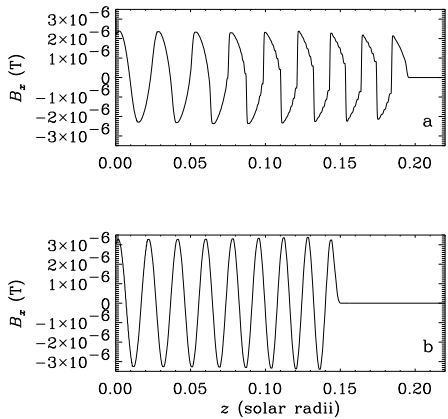}
\caption{ Evolution of $B_x$ for linearly ({\bf a}) and circularly ({\bf b})
polarized Alfv\'{e}n waves propagating through a homogeneous medium. }
\end{figure}

\subsection{Stratified models}

We discus in this paper two models: one with $(\frac{B_x}{B_z})_0 = 0.1$ and one
with $(\frac{B_x}{B_z})_0 = 1.0$, both have a background magnetic field
$B_z=3 \times 10^{-5}$\, T.

At the bottom of the grid $v_{\rm A}$ is smaller than $c_{\rm s}$ and equal 
to $3.8 \times 10^4$ m\,s$^{-1}$, but it increases with $z$ and eventually 
becomes higher than $c_{\rm s}$.
Density is taken to be equal to $5 \times 10^{-13}$\,kg\,m$^{-3}$ at the
bottom of the corona and varies with $z$ as
\begin{eqnarray}
\rho_0(z)=\rho_0(0) \exp(-\frac{R}{H} \frac{z}{R+z})
\end{eqnarray}
\noindent where $H_0= 6.1\,10^7$\,m is the scale height 
at $z=0$, and $R$ is the solar radius.

The wave front of the Alfv\'en wave pushes  matter upwards
forming a density perturbation which behaves as an acoustic precursor as 
long as $v_{\rm A}$ is smaller than $c_{\rm s}$. When $v_{\rm A}$ becomes larger 
than $c_{\rm s}$ the Alfv\'{e}n wave begins to overtake the precursor, which 
however persists and propagates forward with a speed close to the speed of sound
(Fig. 3a).
For a low amplitude Alfv\'{e}n wave the density enhancement almost keeps its
shape as it propagates while for a high amplitude wave the density 
enhancement changes gradually.

\begin{figure}
\epsfxsize=8.8cm
\epsfbox{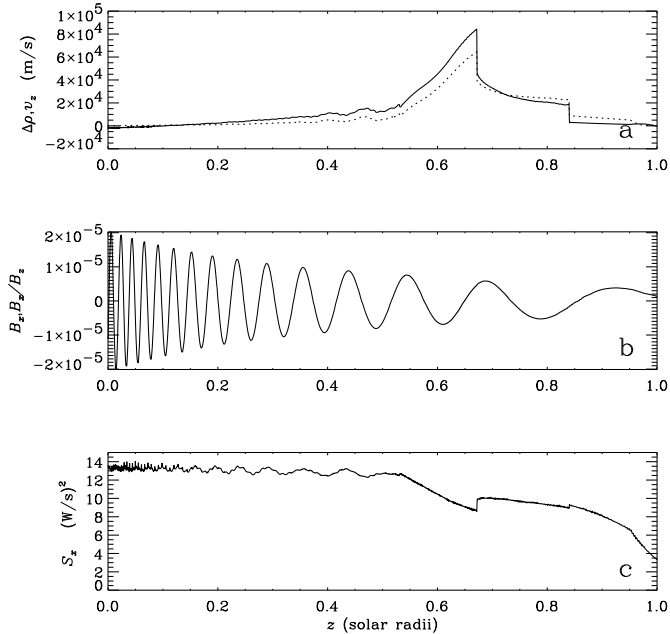}
\caption{\label{p4c}Magnetohydrodynamic waves propagating through a 
stratified
medium with $(\frac{B_x}{B_z})_0 =1$ and $B_z=3 \times 10^{-5}$\,T.
{\bf a} $ \Delta \rho= \rho - \rho_0$ (solid line) and $v_z$
(dashed line) as a function of distance, $z$ at time 4375 s. $\Delta \rho$
has been converted to velocity units by multiplying with $c_{\rm 
s}/{\rho_0}$.
{\bf b} $B_x$ (solid line) as a function of distance $z$ at the same  
time. {\bf c} The $z$-component of the Poynting flux vector 
at the same time.}
\end{figure}

In Fig. 3c we see a decrease in the Poynting flux in front the acoustic 
precursor.
Fig. 4 shows the time average over one Alfv\'{e}n wave period of the
Poynting flux and the integrated work it does via the Lorentz force.
In the linearly polarized case the wave loses $70 \% $ of its Poynting flux 
below $z=0.4$ R$_\odot$
and very little of it is dissipated after that (see Boynton \& Torkelsson  
(1996) Fig. 9) while in
the circularly polarized case the wave loses very little of its Poynting 
flux below $z=0.5$, but it has lost
almost half of it at $z=0.8$ R$_\odot$ by doing mechanical work  on 
the plasma.

\section{Conclusions}

In constructing models of Alfv\'en wave driven stellar winds it is
important to understand the mechanism by which the Alfv\'en waves are
damped.  A linearly polarized Alfv\'en wave of high amplitude can
steepen and form current sheets even in a homogeneous medium, which
leads to a quick damping of the wave.
While a circularly polarized Alfv\'en wave does not dissipate its
energy in a homogeneous medium, we have found that it may do so in
a stratified medium.  The dissipation sets in later than for a 
linearly polarized Alfv\'en wave though, but a significant fraction
of its Poynting flux may still be spent on doing work on the medium
through which the Alfv\'en wave propagates.

\begin{figure}
\epsfxsize=6.0cm
\epsfbox{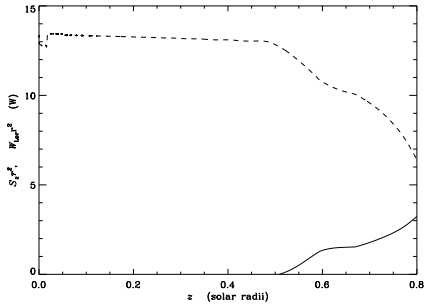}
\caption{ Time average over one Alfv\'{e}n wave period of
the Poynting flux (dashed line) and the integrated work done via the
 Lorentz force (solid line) for a circularly polarized Alfv\'{e}n wave
propagating through a stratified  medium with $(\frac{B_x}{B_z})_0 =1$ .}
\end{figure}

\end{document}